\input harvmac
\noblackbox
\input epsf.tex
\overfullrule=0pt
\def\Title#1#2{\rightline{#1}\ifx\answ\bigans\nopagenumbers\pageno0\vskip1in
\else\pageno1\vskip.8in\fi \centerline{\titlefont #2}\vskip .5in}

\font\ticp=cmcsc10

\font\secfont=cmcsc10

\font\cmss=cmss10 
\font\cmsss=cmss10 at 7pt
\newcount\figno
\figno=0
\def\fig#1#2#3{
\par\begingroup\parindent=0pt\leftskip=1cm\rightskip=1cm\parindent=0pt
\baselineskip=11pt
\global\advance\figno by 1
\midinsert
\epsfxsize=#3
\centerline{\epsfbox{#2}}
\vskip 12pt
{\bf Fig.\ \the\figno: } #1\par
\endinsert\endgroup\par
}
\def\figlabel#1{\xdef#1{\the\figno}}
\def\encadremath#1{\vbox{\hrule\hbox{\vrule\kern8pt\vbox{\kern8pt
\hbox{$\displaystyle #1$}\kern8pt}
\kern8pt\vrule}\hrule}}
                                        
%
%
\baselineskip=18pt plus 2pt minus 2pt

%
\def\CH{{\cal H}}
\def\CN{{\cal N}}

\def\IZ{\relax\ifmmode\mathchoice
{\hbox{\cmss Z\kern-.4em Z}}{\hbox{\cmss Z\kern-.4em Z}}
{\lower.9pt\hbox{\cmsss Z\kern-.4em Z}}
{\lower1.2pt\hbox{\cmsss Z\kern-.4em Z}}\else{\cmss Z\kern-.4em }\fi}
\def\IC{\relax\hbox{$\inbar\kern-.3em{\rm C}$}}
\def\IR{\relax{\rm I\kern-.18em R}}

\def\tht{\tilde{S}}

\def\sot{{\widetilde{SO}(4)}}
\def\so{SO(4)}
\def\bl{{\bar{\l}}}
%

%

\def\s{\sigma}
\def\g{\gamma}
\def\t{\tau}
\def\a{\alpha}
\def\b{\beta}

\def\m{\mu}

\def\p{\pi}

\def\w{\omega}

\def\l{{\lambda}}
\def\O{{\Omega}}

%

\def\1{\relax 1 { \rm \kern-.35em I}}

%
\def\frac#1#2{{#1 \over #2}}

\def\p+{{\partial_+}}

\def\half{{1 \over 2}}


\def\sqr#1#2{{\vcenter{\hrule height.#2pt
            \hbox{\vrule width.#2pt height#1pt \kern#1pt
             \vrule width.#2pt} \hrule height.#2pt}}}
\def\square{\mathchoice\sqr64\sqr64\sqr{4.2}3\sqr33}


\def\[{\left [}
\def\]{\right ]}
\def\({\left (}
\def\){\right )}

\def\TrH#1{ {\raise -.5em
                      \hbox{$\buildrel {\textstyle  {\rm Tr } }\over
{\scriptscriptstyle \CH _ {#1}}$}~}}

\Title{\vbox{\baselineskip12pt
\hbox{\ticp TIFR/TH/02-11}
\hbox{hep-th/0203231}
}}
{\vbox{\centerline {Dp Branes in PP-wave Background} }}
\centerline{{\ticp
Atish Dabholkar and Shahrokh Parvizi\footnote{$^{\dagger}$}
{Address after April 2002: {\it Institute for Studies in Theoretical Physics
and Mathematics, POB 19395-5531, Tehran, Iran}}}}
\vskip.1in
\centerline{\it Department of Theoretical Physics}
\centerline{\it Tata Institute of Fundamental  Research}
\centerline{\it Homi Bhabha Road, Mumbai, India 400005.}
\centerline{Email: atish, parvizi@theory.tifr.res.in}
\vskip .1in
\bigskip
\centerline{ABSTRACT}
\medskip
Dirichlet p-branes in the background of pp waves are constructed using
the massive Green-Schwarz worldsheet action for open strings. These
branes are localized at the origin and only for $p=7, 5, 3$ preserve
half the supersymmetries. The spectrum of the brane theory is analyzed
and is found to be in agreement with the spectrum of the small
fluctuations of the world-volume super Yang-Mills theory in this
background. These branes are expected to correspond to objects that
are nonperturbative in $N$ in the dual gauge theory.
\bigskip
\bigskip
\Date{March 2002}
\vfill\eject
\def\ajou#1&#2(#3){\ \sl#1\bf#2\rm(19#3)}

%
\lref\MaldacenaRE{
J.~Maldacena,
``The large $N$ limit of superconformal field theories and supergravity,''
Adv.\ Theor.\ Math.\ Phys.\  {\bf 2} (1998) 231,
[arXiv:hep-th/9711200].}
\lref\GubserBC{
S.~S.~Gubser, I.~R.~Klebanov and A.~M.~Polyakov,
``Gauge theory correlators from non-critical string theory,''
Phys.\ Lett.\ B {\bf 428}(1998) 105,
[arXiv:hep-th/9802109].}
\lref\WittenQJ{
E.~Witten,
``Anti-de Sitter space and holography,''
Adv.\ Theor.\ Math.\ Phys.\  {\bf 2} (1998) 253,
[arXiv:hep-th/9802150].}
\lref\AharonyTI{
O.~Aharony, S.~S.~Gubser, J.~Maldacena, H.~Ooguri and Y.~Oz,
``Large N field theories, string theory and gravity,''
Phys.\ Rept.\  {\bf 323} (2000) 183,
[arXiv:hep-th/9905111].}
\lref\MetsaevBJ{
R.~R.~Metsaev,
``Type IIB Green-Schwarz superstring in plane wave Ramond-Ramond
background,'' 
[arXiv:hep-th/0112044].}
\lref\GreenWT{
M.~B.~Green and J.~H.~Schwarz,
``Covariant description of superstrings,''
Phys.\ Lett.\ B {\bf 136} (1984) 367.}
\lref\GreenSG{
M.~B.~Green and J.~H.~Schwarz,
``Properties of the covariant formulation of superstring theories,''
Nucl.\ Phys.\ B {\bf 243} (1984) 285.}
\lref\BerensteinJQ{
D.~Berenstein, J.~Maldacena and H.~Nastase,
``Strings in flat space and pp waves from ${\cal N}$ = 4 super Yang Mills,''
[arXiv:hep-th/0202021].}

\lref\Penrose{
R. Penrose, 
``Any space-time has a plane wave as a limit,''
in {\sl Differential geometry and relativity,}
pp. 271-275, Reidel, Dordrecht, 1976.}

\lref\Guven{R. G\"uven, ``Plane wave limits and
T-duality,'' {\sl Phys. Lett.} B482 (2000) 255,
[arXiv: hep-th/0005061].}

\lref\ItzhakiKH{
N.~Itzhaki, I.~R.~Klebanov and S.~Mukhi,
``PP wave limit and enhanced supersymmetry in gauge theories,''
[arXiv:hep-th/0202153].
}
\lref\GomisKM{
J.~Gomis and H.~Ooguri,
``Penrose limit of $N = 1$ gauge theories,''
[arXiv:hep-th/0202157].
}
\lref\ZayasRX{
L.~A.~Zayas and J.~Sonnenschein,
``On Penrose limits and gauge theories,''
[arXiv:hep-th/0202186].
}

\lref\CveticSI{
M.~Cvetic, H.~Lu and C.~N.~Pope,
``M-theory pp-waves, Penrose limits and supernumerary supersymmetries,''
[arXiv:hep-th/0203229].
}

\lref\CveticHI{
M.~Cvetic, H.~Lu and C.~N.~Pope,
``Penrose limits, pp-waves and deformed M2-branes,''
[arXiv:hep-th/0203082.]
}

\lref\DasCW{
S.~R.~Das, C.~Gomez and S.~J.~Rey,
``Penrose limit, spontaneous symmetry breaking and holography in pp-wave 
background,''
[arXiv:hep-th/0203164].
}

\lref\MichelsonWA{
J.~Michelson,
``(Twisted) toroidal compactification of pp-waves,''
[arXiv:hep-th/0203140.]
}

\lref\FloratosUH{
E.~Floratos and A.~Kehagias,
``Penrose limits of orbifolds and orientifolds,''
[arXiv:hep-th/0203134.]
}

\lref\GursoyTX{
U.~Gursoy, C.~Nunez and M.~Schvellinger,
``RG flows from Spin(7), CY 4-fold and HK manifolds to AdS, Penrose  limits and pp waves,''
[arXiv:hep-th/0203124.]
}

\lref\TakayanagiHV{
T.~Takayanagi and S.~Terashima,
``Strings on orbifolded pp-waves,''
[arXiv:hep-th/0203093.]
}

\lref\KimFP{
N.~w.~Kim, A.~Pankiewicz, S.~J.~Rey and S.~Theisen,
``Superstring on pp-wave orbifold from large-N quiver gauge theory,''
[arXiv:hep-th/0203080.]
}

\lref\AlishahihaEV{
M.~Alishahiha and M.~M.~Sheikh-Jabbari,
``The PP-wave limits of orbifolded AdS(5) x S**5,''
[arXiv:hep-th/0203018.]
}

\lref\HatsudaXP{
M.~Hatsuda, K.~Kamimura and M.~Sakaguchi,
``From super-AdS(5) x S**5 algebra to super-pp-wave algebra,''
[arXiv:hep-th/0202190.]
}

\lref\MetsaevRE{
R.~R.~Metsaev and A.~A.~Tseytlin, 
``Exactly solvable model of
superstring in plane wave Ramond-Ramond background,''
[arXiv:hep-th/0202109].
}
\lref\BlauNE{
M.~Blau, J.~Figueroa-O'Farrill, C.~Hull and G.~Papadopoulos,
``A new maximally supersymmetric background of IIB superstring theory,''
JHEP {\bf 0201}, 047 (2002)
[arXiv:hep-th/0110242].
}
\lref\BlauDY{
M.~Blau, J.~Figueroa-O'Farrill, C.~Hull and G.~Papadopoulos,
``Penrose limits and maximal supersymmetry,''
[arXiv:hep-th/0201081].
}
\lref\BlauRG{
M.~Blau, J.~Figueroa-O'Farrill and G.~Papadopoulos,
``Penrose limits, supergravity and brane dynamics,''
[arXiv:hep-th/0202111].
}
\lref\PolchinskiMT{
J.~Polchinski,
``Dirichlet-Branes and Ramond-Ramond Charges,''
Phys.\ Rev.\ Lett.\  {\bf 75}, 4724 (1995)
[arXiv:hep-th/9510017].
}

\lref\McGreevyCW{
J.~McGreevy, L.~Susskind and N.~Toumbas,
``Invasion of the giant gravitons from anti-de Sitter space,''
JHEP {\bf 0006}, 008 (2000)
[arXiv:hep-th/0003075].
}

\lref\BalasubramanianNH{
V.~Balasubramanian, M.~Berkooz, A.~Naqvi and M.~J.~Strassler,
``Giant gravitons in conformal field theory,''
[arXiv:hep-th/0107119].
}

\lref\BerensteinKE{
D.~Berenstein, C.~P.~Herzog and I.~R.~Klebanov,
``Baryon spectra and AdS/CFT correspondence,''
[arXiv:hep-th/0202150].
}

\lref\BilloFF{
M.~Billo' and I.~Pesando,
``Boundary states for GS superstrings in an Hpp wave background,''
[arXiv:hep-th/0203028].
}

\lref\ChuIN{
C.~S.~Chu and P.~M.~Ho,
``Noncommutative D-brane and open string in pp-wave background with  
B-field,''
[arXiv:hep-th/0203186].
}

\newsec{\secfont Introduction}
Recently it has been proposed that a particular subsector of the
four-dimensional $SU(N)$ gauge theory with $\CN=4$ supersymmetry is
dual to Type-IIB superstring theory in a pp wave background
\BerensteinJQ.  This duality is deduced by taking a
scaling limit of the usual correspondence
\refs{\MaldacenaRE,\GubserBC,\WittenQJ, \AharonyTI} 
between $\CN=4$ super Yang-Mills theory and Type-IIB superstring in
$AdS_5\times S^5$ background.  On the Type-IIB side, the scaling limit
is equivalent to the `Penrose limit'
\refs{\Penrose, \Guven} of the geometry near a null geodesic in
$AdS_5\times S^5$ carrying a large angular momentum $J$ on the $S^5$
\refs{\BlauDY,\BerensteinJQ}. In
the gauge theory, this amounts to picking a $U(1)_R$ subgroup of the
$SU(4)$ R-symmetry and focusing on operators with a $U(1)_R$ charge
$J$ and conformal weight $\Delta$ which both scale as $\Delta, J\sim
\sqrt{N}$ keeping the difference $\Delta -J$ finite in the large-N
limit.
The metric of the pp wave is given by
\eqn\ppmetric{
 ds^2 = 2dx^+dx^- + \sum_{I=1}^8 \left( dx^Idx^I - 
\mu^2 x^Ix^I dx^+dx^+ \right),} 
and there is a constant R-R 5-form flux,
\eqn\constf{ F_{+1234}=F_{+5678} = 2 \mu,}
where $\mu$ is a mass parameter characterizing the spacetime.
Type-IIB string theory in this background has a number of remarkable
properties. The background is maximally supersymmetric and thus
preserves $32$ supersymmetries \refs{\BlauNE, \BlauDY}. Moreover, the
world-sheet theory describing the string motion in this background is
exactly solvable. In particular, in the light-cone gauge,$X^+ = p^+
\tau$, the Green-Schwarz action \GreenWT\ contains eight free massive
bosons and fermions \refs{\MetsaevBJ,
\BerensteinJQ, \MetsaevRE} and is given by,
\eqn\action{
S = {1 \over 2 \pi \alpha'}
\int d\tau \int_0^{\pi} d \sigma
\left[ \half {\partial_+ X^I}{\partial_- X^I}
 - \half m^2 (X_I)^2 + i \bar S ( \not \partial + m \Pi) S \right]
 }
where $\partial_{\pm}=\partial_{\tau}\pm\partial_{\sigma}$ on the
worldsheet and $m\equiv \mu p^+$.  Here $X^I, I=1,\ldots, 8$, are the
transverse coordinates that transform as a vector $\bf 8_v$ of the
transverse $SO(8)$ whereas $S$ is a Majorana spinor on the world-sheet
that transforms as a positive chirality spinor $\bf 8_s$ under the
$SO(8)$.  The appearance of $\Pi = \gamma^{1234}$ in the action breaks
the $SO(8)$ symmetry to $SO(4)\times SO(4)$.  In this paper we
investigate the open string sector of this theory. In $\S2$ we find
that Dirichlet p-branes of various orientations are possible and are
localized at the origin of the pp wave. They are supersymmetric only
for $p=3, 5, 7$. In $\S3$ we analyze the spectrum of the worldvolume
theory and find it to be in agreement with the spectrum of low
lying excitations of world-volume super Yang-Mills theory of these
branes in the pp wave background. The Chern-Simons coupling of the
gauge field to the 5-form field is crucial for obtaining this agreement.
We conclude in $\S4$ and discuss the
supersymmetry algebra in the Appendix.

\newsec{\secfont Massive Open Superstrings}
The equations of the string that follow from \action\ are given by
\eqn\eqmotion{\eqalign{
\partial_-\partial_+ X^I + m ^2 X^I = & 0, \cr
\partial_+ S^{1 a} -m  \Pi^a_b S^{2 b} = & 0, \cr
\partial_- S^{2 a} +m  \Pi^a_b S^{1 b} = & 0, \cr
}}
where $a, b$ are $\bf 8_s$ spinor indices and the labels $1, 2$ refer
to the two chiralities on the worldsheet.  These equations are
supplemented by open string boundary conditions. To describe a
Dirichlet p-brane, we impose Neumann boundary conditions on $p-1$
coordinates and Dirichlet boundary conditions on the remaining
transverse coordinates.
\eqn\bbound{\eqalign{
\partial_\sigma X^{r} = & 0 , \;\;  r = 1,\cdots,p-1 \cr  
\partial_\tau X^{r'} = & 0 ,   \;\;  r' = p, \cdots, 8.  \cr
}} 
Notice that the light cone gauge $X^+=p^+\t$ implies that 
${\partial{X^+} \over \partial{\sigma}} =0$. 
Thus, $X^+$ necessarily satisfies
Neumann boundary condition. Moreover, solving the Virasoro
constraints, $X^-$ can be written as
\eqn\virasoro{
\partial_{\sigma} X^- = {-1\over p^+}\partial_{\sigma}X^I \partial_{\t}
X^I + {\rm fermions},} 
which implies that $X^{-}$ too necessarily satisfies
Neumann boundary condition.
For the fermionic coordinates, the boundary condition is
\eqn\fbound{
S^1|_{\sigma=0,\pi}=\Omega S^2|_{\sigma=0,\pi},} 
where, as in flat space, $\Omega$ is a matrix $\Pi_k \g^k$ acting the
$\bf 8_s$ representation with the product running over the index $k$
that labels the Dirichlet directions. We are interested in BPS objects
that preserve sixteen supesymmetries. The choice of allowed $\Omega$
is constrained further by the following two conditions:
\eqna\eIIi{$$\eqalignno{
[\O, \g] & =0&\eIIi a\cr
\O \Pi \O \Pi & = -1&\eIIi b\cr}$$}
In the first condition \eIIi{a}, $\gamma$ is the chirality matrix of
$SO(8)$ and it follows from the fact that both fermions have the same
$SO(8)$ chirality in Type-IIB string. This condition allows only odd
$p$ branes in the theory as in flat space. It is easy to see from the
equations of motion for the fermions that, because of the mass term,
eight fermion zero modes exist only if condition \eIIi{b} is
satisfied. Note that the `zero' modes have zero worldsheet momentum
($n=0$), but nonzero worldsheet energy because of the mass term. They
satisfy a Clifford-like algebra and generate the supermultiplet of the
unbroken supersymmetries.  The allowed choices for $\O$ consistent
with these two constraints are
\eqn\allowed{\eqalign{
{\rm D7}:&\;\; \g^i \g^j , \;\; \g^{i'} \g^{j'}, \cr
{\rm D5}:&\;\; \g^{i'} \g^i \g^j \g^k, \;\; \g^i \g^{i'} \g^{j'}
\g^{k'}, \cr
{\rm D3}:& \;\;\g^{i'} \g^{j'} \Pi, \;\; \g^i \g^j \Pi,\cr}} where $i, j,
k= 1, \cdots, 4$ and $i', j', k'= 5, \cdots, 8$. 
Each choice for $\O$, corresponds to a Dp-brane transverse to the
indices appearing in the above $\g$ matrices. Because the rotation
symmetry is reduced to $SO(4)\times SO(4)$, different orientations of
branes are physically distinct. Note that $D1$ and $D9$ branes are not
allowed.  The mode expansion of the bosonic coordinates satisfying the
boundary conditions and the equations of motion is given by
\eqn\bexpan{\eqalign{
X^{r}(\sigma,\tau) =&  x_0^r \cos m\tau +{1 \over m} p_0^r \sin
m\tau
+i \sum_{n\neq 0} {1 \over \omega_n} \alpha^r_n {\rm
e}^{-i\omega_n\tau}\cos n\sigma \cr
X^{r'}(\sigma,\tau) =&   
 \sum_{n\neq 0} {1 \over \omega_n} \alpha^{r'}_n {\rm
e}^{-i\omega_n\tau}\sin n\sigma 
,\cr}}
with
\eqn\freq{
\w_n = {\rm sgn}(n) \sqrt{n^2 + m^2}.
}
In the mode expansion of the Dirichlet directions $X^{r'}$, there is
no zero mode consistent with the supersymmetry algebra corresponding
to the position of the brane. As a result, the Dp brane is stuck at
the origin $X^{r'}=0$. The zero mode of the Neumann directions can
fluctuate but experiences a harmonic oscillator potential centered at
the origin.  The mode expansion of the fermions is
\eqn\fermions{\eqalign{
S^{1} =& S_0 \cos m\tau + \tht_0 \sin m\t + \sum_{n\neq 0}
c_n\left({i \over m}(\w_n-n) \Pi S_n \phi_n + 
\tilde{S}_n {\tilde\phi}_n
\right) \cr
S^{2} =& -\Pi S_0 \sin m\tau + \Pi \tht_0 \cos m\t +
\sum_{n\neq 0} c_n
\left(S_n \phi_n -{i\over m}(\w_n-n) \Pi \tilde{S}_n 
{\tilde\phi}_n \right),\cr}}
where,
\eqn\coef{\eqalign{
\phi_n = & \exp{-i(\w_n \tau + n\sigma)}, \qquad {\tilde\phi}_n =
\exp{-i(\w_n\tau -n\sigma)}\cr
& c_n = \left( 1+{(\w_n - n)^2 \over m^2} \right)^{-1/2},
}}
The boundary condition on fermions with a general $\O$
subject to conditions \eIIi\ gives us,
\eqn\solution{\eqalign{
\tht_0 =& -\O \Pi S_0 \cr
\tilde{S}_n =& \O S_n, \;\;\;\; n \neq 0
.}}
For canonical quantization we introduce canonical momenta
\eqn\momentum{\eqalign{
P^I = \dot{X}^I,& \;\;\;\; I=1,\cdots, 8 \cr {\cal P}^{{\cal I} a} =
S^{{\cal I} a},& \;\;\;\; {\cal I}=1, 2.}}  
The phase space for the fermions is constrained because the momenta in
\momentum\ are proportional to the coordinates. Also a further constraints
come from the boundary conditions \solution. As usual, these
constraints can be consistently incorporated by using Dirac Brackets.
We regard the $S_n$ oscillators as independent variables and solve for
$\tht_n$ using \solution. The Dirac brackets for the independent
variables are then given by,
\eqn\brackets{\eqalign{
\{\a_n^I, \a_l^J\}_{DB} =& {i\w_n \over 2} \delta_{n+l,0} \delta^{IJ}\cr
\{S_0^{a}, S_{0 b}\}_{DB} =& {i \over 4}\delta^{a}_{b} \cr
\{S_n^{a}, S_{l b}\}_{DB} =& {i \over 4}\delta^{a}_{b} 
\delta_{n+l,0}.\cr}}
Canonical quantization proceeds by replacing the classical brackets
with (anti-)commutators including a factor of $i$. We rescale the
oscillators properly to find
\eqn\commut{\eqalign{
[\bar{a}_n^I, a_l^J] = \delta_{n+l,0} \delta^{IJ},& \;\;\;\; [\bar{a}_0^i,
a_0^j] = \delta^{IJ}, \cr
\{S_0^{a}, S_{0 b}\} = {1 \over 4}\delta^{a}_{b},& \;\;\;\; \{S_n^{a},
S_{lb}\} = {1 \over 4}\delta_{n+l,0}\delta^{a}_{b} ,}}
in which
\eqn\modes{
a_0^r = {1 \over \sqrt{2m}}(p_0^r +i m x_0^r), \;\;\;\;\;
\bar{a}_0^{r} = {1 \over \sqrt{2 m}}(p_0^r -i  m x_0^r), \;\;\;\;\;
a_n^I = \sqrt{2\over |\w_n|} \a_n^I.} 
Note that for the allowed choices of $p=3, 5, 7$, there are eight
independent real fermionic zero modes. They are proportional to the
supercharges and satisfy a Clifford-like algebra. As we describe in
$\S3$, the representation of this algebra gives a short
$16$-dimensional supermultiplet.  The Hamiltonian can now be written
as
\eqn\hamilton{
H={1 \over \pi p^+}\int_0^{\pi} \left( {1\over 2}( {\cal P}_I^2 +
{X'}_I^2 + \m^2 X_I^2 ) + i (S^1 \dot{S}^1 +S^2 \dot{S}^2) \right)d\s
,}
in which we have used the equations of motion for fermions. In terms
of the quantized oscillators, the Hamiltonian takes the form
\eqn\hamilmode{\eqalign{
H =& E_0 +E_{\cal N}, \cr
E_0=& {m \over p^+} \left(\sum_{r=1}^{p-1}\bar{a}_0^r a_0^r - 
2 i S_0 \O \Pi
S_0 +  e_0\right), \cr
E_{\cal N}=& {1 \over  p^+} \left(\sum_{n\neq 0} a^I_n a^I_{-n} +
i\sum_{n\neq 0} \w_n S_n S_{-n} \right).}}
The zero point energy $e_0={p-1 \over 2}$ is a result of normal
ordering $p-1$ harmonic oscillator bosonic zero modes.
In the next section we consider $p=7$ and describe in detail the
worldvolume spectrum of the D7-brane. 
\newsec{\secfont World-volume Spectrum}
Let us consider a D7-brane extending along the $(+-123456)$ directions
so that $\O=\g^{78}$. The $SO(4) \times \sot $ symmetry of the pp wave
is broken by the boundary conditions to the longitudinal $\so \times
SO(2)$ times a transverse $SO(2)'$.  Writing $SO(4)$ factor as
$SU(2)_L\times SU(2)_R$, we have the embedding
\eqn\group{\eqalign{
SO(8) & \supset  SU(2)_L\times SU(2)_R \times 
\widetilde{SU}_L(2)\times \widetilde{SU}(2)_R \cr
 & \supset SU(2)_L\times SU(2)_R \times SO(2) \times SO(2)'.\cr}}
The $SO(2)\times SO(2)'$ are rotations in the $56$ and $78$ and are
generated by $T_{56}$ and $T_{78}$ respectively. In terms of the
generators of $\widetilde{SU}(2)_L\times
\widetilde{SU}(2)_R$, they are given by
\eqn\soso{\eqalign{
T_{56} =& \tilde{J}_{3L}+\tilde{J}_{3R} \cr
T_{78} =& \tilde{J}_{3L}-\tilde{J}_{3R}. \cr}}
Under this embedding the spinor decomposes as 
\eqn\spinor{
\bf{8_s} \sim ({\bf 2}, {\bf 1})^{(\half, \half)} \oplus
({\bf {\bar 2}}, {\bf 1})^{(-\half, -\half)} \oplus 
({\bf 1}, {\bf 2})^{(\half, -\half)} \oplus 
({\bf 1}, {\bf {\bar 2}})^{(-\half, \half)},}
where the superscripts denote $SO(2)\times SO(2)'$ charges. 
Using this decomposition, we can now organize the zero modes of the
spinor in terms of fermionic creation and annihilation operators:
\eqn\fermiond{\eqalign{
\bar{\l}_{\a} \equiv S_{0\a}^{(\half, \half)},& \;\;\;\;  
{\l}_{\a} \equiv S_{0 \a}^{(-\half, -\half)}, \cr
\bar{\l}_{\dot{\a}} \equiv S_{0 \dot{\a}}^{(-\half, \half)},&  \;\;\;\;
{\l}_{\dot{\a}} \equiv S_{0 \dot{\a}}^{(\half, -\half)}.}}
where $\a$ and $\dot{\a}$ are the doublet indices of $SU(2)_L$ and
$SU(2)_R$ respectively.
Let us now look at the Hamiltonian \hamilmode\ for the fermionic zero
modes. The matrix $\O\Pi$ is nothing but $\g^{56}$, which is the
generator of $T_{56}$ in the spinor representation. We have chosen the
basis \fermiond\ such that the Hamiltonian is diagonal and can be
written as
\eqn\hamildiag{
E_0 = \mu \left( {\rm bosons} + \bar{\l}_{\a} \l^{\a} -
\bar{\l}_{\dot{\a}} \l^{\dot{\a}} +e_0 \right)}
with zero-point energy $e_0 = 3$.  The commutation relations in this
basis are given by
\eqn\ocsil{\eqalign{
\{\bar{\l}_{\a}, \l^{\b}\} =& \delta_{\a}^{\b}, \cr
\{\bar{\l}_{\dot{\a}}, \l^{\dot{\b}}\} =& \delta_{\dot{\a}}^{\dot{\b}}.}}
We choose the Fock vacuum that is $SO(4)\times SO(2)$ invariant with
$T_{78}$ charge $-1$:
\eqn\vacuum{
 a_0 |0, -1\rangle = 0, \;\;\;\;\; {\l}^{\a} |0, -1 \rangle = 0,
\;\;\;\;\; {\l}^{\dot{\a}} |0, -1 \rangle = 0.}
The labels of the ket denote $T_{56}$ and $T_{78}$ charges
respectively. The low lying multiplet of ground states is constructed
on top of this vacuum by successively acting with the zero-mode
creation operators $\bl_{\a}$ and $\bl_{\dot{\a}}$.  This ground state
multiplet is summarized in the table below.  The entire Hilbert space
of the 7-brane world volume theory is obtained by the action of
various creation operators of the bosonic and fermionic modes on the
ground state multiplet. Note that as described in the Appendix, the
supersymmetry generators in this background do not commute with the
light-cone Hamiltonian. However, the entire superalgebra closes.
Thus the supersymmetry algebra is not a symmetry of the Hamiltonian but
rather a spectrum-generating algebra.
$$\vbox{\settabs 6\columns
\+State && Representation && Energy & Field \cr
\+$|0, -1\rangle$ &&  \quad $({\bf 1}, {\bf 1})^{(0, -1)}$
&& \quad 3 & \quad $\phi$ \cr
\+$\bl_{\a}|0, -1\rangle$ 
&& \quad $({\bf 2}, {\bf 1})^{(\half, -\half)}$ && \quad $3 + 1$ 
& \quad ${\psi}_{\a}$ \cr
\+$\bl_{\dot{\a}}|0, -1\rangle$ 
&& \quad $({\bf 1}, {\bf 2})^{(-\half, -\half)}$ && \quad $3- 1$
& \quad ${\psi}_{\dot{\a}}$ \cr
\+$\bl_{\a}\bl_{\b}|0, -1\rangle$ 
&& \quad $({\bf 1}, {\bf 1})^{(1, 0)}$ && \quad $3+2$
& \quad $A$ \cr
\+$\bl_{\a} \bl_{\dot{\a}} |0, -1\rangle$ 
&& \quad $({\bf 2}, {\bf 2})^{(0, 0)}$ && \quad $3$
& \quad $A_i$ \cr
\+$\bl_{\dot{\a}} \bl_{\dot{\b}}|0, -1\rangle$ 
&& \quad $({\bf 1}, {\bf 1})^{(-1, 0)}$ && \quad $3 - 2$
& \quad $\bar{A}$ \cr
\+$\bl_{\a}\bl_{\dot{\a}} \bl_{\dot{\b}}|0, -1\rangle$ 
&& \quad $({\bf 2}, {\bf 1})^{(-\half, \half)}$ && \quad $3-1$
& \quad ${\bar\psi}_{{\a}}$ \cr
\+$\bl_{\a}\bl_{\b}\bl_{\dot{\a}} |0, -1\rangle$ 
&& \quad $({\bf 1}, {\bf 2})^{(\half, \half)}$ && \quad $3+ 1 $  
& \quad ${\bar\psi}_{\dot{\a}}$ \cr
\+$\bl_{\a}\bl_{\b}\bl_{\dot{\a}} \bl_{\dot{\b}}|0, -1\rangle$ 
&& \quad $({\bf 1}, {\bf 1})^{(0, 1)}$ && \quad 3 & \quad $\bar{\phi}$
\cr }$$ 
For a given open string state in the first column, its $SU(2)_L\times
SU(2)_R \times SO(2)\times SO(2)'$ representation is listed in the
second column and the energy in units of $\mu$ is listed in third
column. The last column lists the corresponding mode of the
world-brane supergauge field. The entries in this column follow from
an analysis of the spectrum of small fluctuations of the world brane
theory which we now describe. As we will see, the Chern-Simons
coupling of the gauge field to the five-form field plays a crucial
role.

The low effective theory on a 7-brane contains a gauge field
$A_M$, a complex scalar $\phi$ and gauginos. The scalar field does not
couple to the five-form field and hence the small fluctuation of the
scalar field is a solution of the equation
\eqn\scalar{
\square \phi = 0, \qquad  \square \equiv 
{1\over \sqrt{-g} } \partial_M (\sqrt{-g} g^{MN} \partial_N )
= 2\partial{+}\partial{-} + \mu^2 x_I^2\partial_{-}^2 +\partial_I^2,}
where $M= +, -, 1, 2, 3, 4, 5, 6$ and $I=1, \ldots, 6$ are the
worldvolume indices.  Let us now Fourier transform in $x^-, x^I$
\eqn\fourier{
\phi(x^+,x^-,x^I) =\int
{dp^+d^{8}p} \, e^{{\rm i }(p^+x^- + p^Ix^I)}\, 
\tilde \phi(x^+,p^+,p^I).}
Then the equation \scalar\ becomes
\eqn\ham{
i\partial_{+} \tilde{\phi} = H \phi \equiv  
\frac{1}{2p^+}(p_I^2 - m^2   \partial_{p^I}^2 ) \phi.}
Introducing the standard creation and annihilation operators, the
normal ordered Hamiltonian become
\eqn\hamtwo{
H = \mu (\bar{a}^I a^I +3).}
Note that the energy of the Fock space ground state $| 0\rangle$
annihilated by the creation operators is $3$ because there are $6$
transverse directions on the 7-brane worldvolume. The gauge field
dynamics is governed by the Yang-Mills action in the pp-wave
background:
\eqn\gaugeaction{
S_B = \int_{\Sigma} \half F\wedge * F + C^{(4)}\wedge F\wedge F}
where $C^{(4)}$ is the 4-form potential of the background 
Ramond-Ramond gauge field. 
The equation of motion that follows from this action is
\eqn\gaugemotion{
d* F = F^{(5)} \wedge F.}
In the light-cone gauge,  $A_{-}=0$, the component $A_{+}$ is determined
in terms of the physical transverse modes $A_r, r=1,\ldots 6$.  To
find the gauge field fluctuations we note that, under the worldvolume
rotation symmetry $SO(4)\times SO(2)$, $A_m$ decomposes as $A_i ({\bf
4}^0) \oplus A ({\bf 1}^1)\oplus {\bar A} ({\bf 1}^{-1})$.  The CS
coupling gives a term proportional to $\mu \epsilon^{+-1234pq}$.  The
$\epsilon$ tensor effectively acts like the rotation matrix $T_{56}$
in the $56$ plane and as a result the transverse modes $A_m$ satisfy
the equation
\eqn\motionwo{
(\square + 4 i \mu T_{56} \partial_- ) A_r =0.}
The normal ordered Hamiltonian now has an extra piece compared to
\hamtwo\ and is given by
\eqn\hamthree{
H = \mu (\bar{a}^I a^I +3 + 2 T_{56}).}
This explains the splittings listed in column four above. The $A_i$
components are neutral under $T_{56}$ and hence their energy is $3$
whereas the combinations $A_5 \pm i A_6$ have energy $3\pm 2$.  The
equations of motion for the gaugino are related to \gaugemotion\ by
supersymmetry and also contain a term proportional to the
5-form field. The second order equation for the gaugino that follows
leads to the same Hamiltonian as in \hamthree. Now, under the above
decompositions all fermions have charge $\pm \half$ under $T_{56}$ and
hence their energies are given by $3\pm 1$. Thus the energies and
representations of all low-lying modes of the worldvolume theory are
in agreement with the quantized spectrum of the massive worldsheet
theory.

\newsec{\secfont Conclusions}

Open strings usually describe nonperturbative Dirichlet p-branes in
the theory \PolchinskiMT. We have seen that a variety of Dp branes are
possible in the pp wave background. As explained earlier, the light
cone directions of the open string are always Neumann. Thus, one leg
of the brane always wraps the direction in $S^5$ used in the Penrose
limit.  In this respect, these D-branes resemble the giant gravitons
in that they are brane configurations that wrap the $S^5$ and preserve
half the supersymmetries \McGreevyCW. We expect that these branes
correspond to some objects in the gauge theory that are
nonperturbative in $N$ such as the baryons \refs{\BalasubramanianNH,
\BerensteinKE}. It would be interesting to find 
precise gauge theory interpretation of these states. 
There are a number of interesting papers studying various aspects of 
the pp-wave background 
\refs{\ItzhakiKH, \GomisKM, \ZayasRX, \HatsudaXP, 
\AlishahihaEV, \KimFP, \TakayanagiHV, \GursoyTX, 
\FloratosUH, \DasCW, \MichelsonWA, \CveticHI, \CveticSI}. 
Considerations in this paper would be relevant in these other contexts
also.

There are a number of open questions about the D-branes such as their
interactions with the closed string modes which we hope to return to
in future. At present, various aspects that are well-known about the
massless theory such as the open-closed channel duality, determination
of the critical dimension, incorporation of interactions etc. are not
fully understood for the massive theory and are in need of further
elucidation.

{\it Note Added:} In a related work \BilloFF\ that appeared during the
course of this investigation, various allowed boundary states in the
closed string sector of the theory have been constructed. These
boundary states would naively correspond, in the open string channel,
to Dirichlet boundary conditions for the light-cone coordinates.
Thus, they do not seem to be directly related to the p-branes
considered in this paper which are Neumann in the lightcone
coordinates. Noncommutative Dp branes in the pp-wave background have
been considered in \ChuIN\ with some overlap with this paper.

\medskip
\leftline{ \secfont Acknowledgments}
\medskip
We would like to thank Sunil Mukhi and Sandip Trivedi and especially
Anindya Biswas for useful discussions and M. Sheikh-Jabbari for
comments on the draft. A. D. would like to acknowledge the hospitality
of the theory group at SLAC and Stanford University where part of this
work was completed.
\vfill\eject
\medskip
\leftline{\secfont Appendix: Supersymmetry Algebra}
\medskip

We will now describe the supersymmetry algebra and exhibit the supersymmetry
of the spectrum. We will use the Super-Noether charges for the closed string
introduced by Metsaev
\MetsaevBJ :
\eqn\noether{\eqalign{
{\bf P}^+ = p^+ ,& \;\;\;\;\;\; {\bf P}^I={1 \over \pi}\int d\s(P^I
\cos{m\t} +
X^I m \sin{m\t} ) \cr
{\bf J}^{+I}=&{i p^+ \over \pi} \int d\s({1 \over m} P^I \sin{m\t}
- X^I \cos{m\t})
\cr
Q^+ =& {\sqrt{2 p^+}\over \pi} \int d\s {\rm e}^{i m\t \Pi} (S^1 +i
S^2) ,
\cr 
\bar{Q}^+ =& {\sqrt{2 p^+}\over\pi} \int d\s {\rm e}^{-i m\t \Pi} (S^1 -
i
S^2)  \cr
{\bf J}^{ij} =& {1 \over \pi}\int d\s \left(x^i P^j -x^j P^i - i (S^1 - i
S^2) \g^{ij}(S^1 +i S^2)\right), \cr
{\bf J}^{i'j'} =&{1 \over \pi} \int d\s \left(x^{i'} P^{j'} -x^{j'} P^{i'}
- i (S^1 - iS^2) \g^{i'j'}(S^1 + i S^2)\right), \cr
Q^{-1} =& {2 \over \pi\sqrt{p^+}} \int d\s [(P^I - \acute{X}^I)\g^I S^1 -
m X^I \g^I \Pi S^2], \cr
Q^{-2} =& {2 \over \pi\sqrt{p^+}} \int d\s [(P^I - \acute{X}^I)\g^I S^2 +
m X^I \g^I \Pi S^1], \cr
{\bf P}^- =& - H,
}}
in which $(I, J=1,\cdots,8)$, $(i,j=1,\cdots,4)$, $(i',j'=5,\cdots,8)$,
and $Q^-, \bar{Q}^- = (Q^{-1}\pm i Q^{-2})/\sqrt{2}$. 
We use the explicit solutions \bexpan\ and \fermions\ to derive the
supercharges in terms of open string modes. To do this, one can use a
proper doubling of the interval $[0, \pi]$ to $[0, 2\pi]$, such that
all the classical solutions satisfy the open string boundary conditions
for interval $[0, \pi]$ and periodic boundary conditions for $[0,
2\pi]$. The resulting charges will be as follows,
\eqn\pcharges{
{\bf P}^+ = p^+ , \;\;\;\;\;\; {\bf P}^r = p_0^r, \;\;\;\;\;\; 
{\bf P}^{r'} = 0,
}
\eqn\jplus{
{\bf J}^{+r} = -i p^+ x_0^r, \;\;\;\;\;\; {\bf J}^{+r'} = 0,
}
\eqn\jrs{\eqalign{
J^{rs} = a_0^r \bar{a}_0^s - a_0^s \bar{a}_0^r + \half S_0 \g^{rs} S_0 
 + \sum_{n\neq 0} \left[a_n^r a_{-n}^s - a_{n}^s a_{-n}^r + {1\over 4}
S_{-n} \g^{rs} S_n \right],
}}
\eqn\jrrss{\eqalign{
J^{r's'} =  \half S_0 \g^{r's'} S_0 
 +\sum_{n\neq 0} \left[a_n^{r'} a_{-n}^{s'} - a_{n}^{s'} a_{-n}^{r'}
+ {1 \over 4} S_{-n} \g^{r's'} S_n \right]
}}
\eqn\jrrs{
J^{r's} = 0  
}
\eqn\qplus{\eqalign{  
Q^+ = \sqrt{p^+} (1+ i \O^T) S_0, \;\;\;\;\;\; 
\bar{Q}^+ = \sqrt{p^+} (1- i \O^T) S_0 
}}
\eqn\qminusone{\eqalign{
\sqrt{2 p^+} Q^{-1} = & 2p_0^I \g^I S_0 - 2m x_0^r\g^r \Pi \O^T S_0 \cr
 & + \sum_{n=1}^{\infty}\left[ \sqrt{2\w_n} c_n a_n^I \g^I \O S_n + {im 
\over\sqrt{2\w_n} c_n}(a_n^r\g^r -a_n^{r'}\g^{r'})\Pi S_n +h.c. \right] \cr
}}
\eqn\qminustwo{\eqalign{
\sqrt{2 p^+} Q^{-2} =& 2p_0^I \g^I \O^T S_0 +2m x_0^r\g^r \Pi S_0 \cr
 & + \sum_{n=1}^{\infty}\left[ \sqrt{2\w_n} c_n (a_n^r\g^r
-a_n^{r'}\g^{r'}) S_n - {im \over \sqrt{2\w_n} c_n} a_n^I\g^I
\Pi \O S_n +h.c. \right]. \cr
}}
We have also used the following identities in the expressions for rotation
generators, $J^{IJ}$,
\eqn\identities{\eqalign{
\g^{rs}+\O\g^{rs}\O^T =& 2 \g^{rs}, \cr
\g^{r's'}+\O\g^{r's'}\O^T =& 2 \g^{r's'}, \cr
\g^{r's}+\O\g^{r's}\O^T =& 0, }} 
Note that in
\qplus, the components $Q^+_{\a}$ and $\bar{Q}^+_{\b}$ are not
independent and as a result half the linear combinations of the
supercharges are zero. This means that the D-brane breaks half the
supersymmetry as expected. To see the superalgebra of the unbroken
supersymmetries explicitly, consider the following combinations,
\eqn\spinors{\eqalign{
(Q^+ + \bar{Q}^+) + i \O (Q^+ - \bar{Q}^+) =& 0 \cr 
Q^{-1} - \O Q^{-2} =& 0. 
}}
Then we can use the following independent supercharges, 
\eqn\consspinors{\eqalign{
q^+ =& \half ((Q^+ + \bar{Q}^+) - i \O (Q^+ - \bar{Q}^+)) = 2 \sqrt{p^+}
S_0 \cr
q^- =& {1\over 2\sqrt{2}}(Q^{-1} + \O Q^{-2}) = {1\over \sqrt{2}} Q^{-1}.   
}}
The supersymmetry algebra the becomes
\eqn\ppjj{\eqalign{
[P^-, P^I] =& \mu^2 J^{+I},  \;\;\;\;\;\;\;
[P^I, J^{+J}] = - \delta^{IJ} P^+,  \;\;\;\;\;\;\;
[P^-, J^{+I}] = P^I, \cr
[P^r, J^{st}] =& \delta^{sr} P^t - \delta^{tr} P^s,  \;\;\;\;\;\;\;
[J^{+r}, J^{st}] = \delta^{sr} J^{+t} - \delta^{tr} J^{+s}, \cr
[J^{rs}, J^{tu}] =& \delta^{st} J^{ru} + {\rm 3 \;\; terms},
\;\;\;\;\;\;\;
[J^{r's'}, J^{t'u'}] = \delta^{s't'} J^{r'u'} + {\rm 3 \;\; terms}, 
}}
\eqn\pjq{\eqalign{
[J^{IJ}, q_{\a}^{\pm}] =& \half q_{\b}^{\pm} (\g^{IJ})^{\b}_{\a}, \cr
[J^{+I}, q_{\a}^{-}] =& \half q_{\b}^{+} (\g^{+I})^{\b}_{\a}, 
\qquad [P^{-}, q_{\a}^{-}] = 0\cr
[P^{r}, q_{\a}^{-}] =& \half \mu q_{\b}^{+} 
(\Pi\O^T\g^{+I})^{\b}_{\a}, \;\;\;\;\;\; 
[P^{-}, q_{\a}^{+}] = \mu  q_{\b}^{+} (\Pi\O^T)^{\b}_{\a},\cr }}
\eqn\antiqq{\eqalign{
\{q^+_{\a}, q^{+}_{\b}\} =& 2 p^+ \delta_{\a\b} \cr
\{q_{\a}^+, q_{\b}^{-}\} =& (\g^{r})_{\a\b} P^r 
+ i \mu (\g^r\Pi)_{\a\b} J^{+r}   \cr
\{q_{\a}^-, q_{\b}^{-}\} =& \delta_{\a\b} P^- 
+ \mu (\g^{rs}\Pi)_{\a\b} J^{rs}.}}
Thus, the supersymmetry algebra closes. The worldvolume states are
organized in a representation of this superalgebra.  Note that in the
limit $\mu=0$ we recover the flat space superalgebra.

\bigskip
\vfill
\eject
\listrefs
\end